\documentclass[sigconf]{acmart}

\AtBeginDocument{%
  \providecommand\BibTeX{{%
    \normalfont B\kern-0.5em{\scshape i\kern-0.25em b}\kern-0.8em\TeX}}}

\setcopyright{acmcopyright}
\copyrightyear{2021}
\acmYear{2021}
\acmDOI{xx.yyyy/zzzzzzzzzzz}

\acmConference[ASHES '21]{ASHES '21: Workshop on Attacks and Solutions in Hardware Security}{Nov. 19, 2021}{Seoul, South Korea}
\acmBooktitle{ASHES '21: Workshop on Attacks and Solutions in Hardware Security, Nov. 19, 2021, Seoul, South Korea}
\acmPrice{15.00}
\acmISBN{978-1-4503-XXXX-X/18/06}

\usepackage{algorithm}
\usepackage{algorithmic}
\usepackage{multirow}
\usepackage{multicol}
\usepackage[flushleft]{threeparttable}
\usepackage{float}

\setcopyright{none}

\begin{document}
\newcommand{\Z}{\mathbb Z}

\title{Design Space Exploration of SABER in 65nm ASIC}


 \author{Malik Imran}
 \affiliation{%
   \institution{Tallinn University of Technology}
   \streetaddress{1 Th{\o}rv{\"a}ld Circle}
   \city{Tallinn}
   \country{Estonia}}
 \email{malik.imran@taltech.ee}
 
  \author{Felipe Almeida}
 \affiliation{%
   \institution{Tallinn University of Technology}
   \streetaddress{1 Th{\o}rv{\"a}ld Circle}
   \city{Tallinn}
   \country{Estonia}}
 \email{felipe.almeida@taltech.ee}

 \author{Jaan Raik}
 \affiliation{%
   \institution{Tallinn University of Technology}
   \streetaddress{1 Th{\o}rv{\"a}ld Circle}
   \city{Tallinn}
   \country{Estonia}}
 \email{jaan.raik@taltech.ee}
 
  \author{Andrea Basso}
 \affiliation{%
   \institution{University of Birmingham}
   \streetaddress{1 Th{\o}rv{\"a}ld Circle}
   \city{Birmingham}
   \country{UK}}
 \email{a.basso@pgr.bham.ac.uk}
 
 \author{Sujoy Sinha Roy}
 \affiliation{%
   \institution{Graz University of Technology}
   \streetaddress{1 Th{\o}rv{\"a}ld Circle}
   \city{Graz}
   \country{Austria}}
 \email{sujoy.sinharoy@iaik.tugraz.at}
 
 \author{Samuel Pagliarini}
 \affiliation{%
   \institution{Tallinn University of Technology}
   \streetaddress{1 Th{\o}rv{\"a}ld Circle}
   \city{Tallinn}
   \country{Estonia}}
 \email{samuel.pagliarini@taltech.ee}




\begin{abstract}
This paper presents a design space exploration for SABER, one of the finalists in NIST's quantum-resistant public-key cryptographic standardization effort. Our design space exploration targets a 65nm ASIC platform and has resulted in the evaluation of 6 different architectures. Our exploration is initiated by setting a baseline architecture which is ported from FPGA. In order to improve the clock frequency (the primary goal in our exploration), we have employed several optimizations: (i) use of compiled memories in a `smart synthesis' fashion, (ii) pipelining, and (iii) logic sharing between SABER building blocks. The most optimized architecture utilizes four register files, achieves a remarkable clock frequency of $1GHz$ while only requiring an area of 0.314$mm^2$. Moreover, physical synthesis is carried out for this architecture and a tapeout-ready layout is presented. The estimated dynamic power consumption of the high-frequency architecture is approximately 184mW for key generation and 187mW for encapsulation or decapsulation operations. These results strongly suggest that our optimized accelerator architecture is well suited for high-speed cryptographic applications.
\end{abstract}

\ccsdesc[100]{Hardware~Application specific integrated circuits}
\ccsdesc[100]{Security and privacy~Hardware security implementation}
\ccsdesc[500]{Security and privacy~Cryptography}

\keywords{SABER, Lattice cryptography, MLWR, Crypto core, ASIC}


\maketitle

\section{Introduction} \label{sec:introduction}
Currently deployed public-key cryptographic schemes, i.e., Rivest Shammir Adleman (RSA) and Elliptic-curve Cryptography (ECC), have their security strength built on the hardness of solving hard mathematical problems such as prime factorization and discrete logarithms. While these crypto schemes have been standardized and, to a large extent, remain useful, the recent advances in the field of quantum computers now threat to break them \cite{SHOR_ALGO}. Therefore, researchers are focusing on designing and investigating quantum-resistant public-key algorithms and protocols to keep future communications secure.

Recently, a competition has been started by the National Institute of Standards and Technology (NIST) for the standardization of post-quantum cryptographic (PQC) public-key protocols \cite{NIST}, i.e., protocols that would not be vulnerable to quantum computers. As the competition approaches its end, the majority of the remaining candidates are based on computationally infeasible lattice problems. One such candidate is a key encapsulation mechanism (KEM) named SABER \cite{SABER}, which is the central piece of this study.

Throughout the standardization/competition process, NIST has considered the security strength of PQC KEM protocols. C/C++ reference implementations of the finalist protocols are available from \cite{NIST}. Naturally, as with ECC and RSA, having accelerators for PQC candidates is of interest as dedicated hardware can achieve significant speed-ups in performance. Examples of hardware accelerators for NIST PQC protocols are presented in \cite{Sinha_Roy_Basso_2020, Three_Algos_2019,Roy_DAC_2021, SAPPHIRE_2019, SABER_Karatsuba_2020,SC_Resistant_Implementation_of_SABER,Lightweight_SABER} where both field programmable gate array (FPGA) and application specific integrated circuit (ASIC) platforms are targeted.

Comparatively, state-of-the-art hardware implementations of SABER \cite{Sinha_Roy_Basso_2020, SABER_Karatsuba_2020} provide significant performance improvements in terms of computational time for the key generation (KeyGen), encapsulation (Encaps) and decapsulation (Decaps) operations. The required computation time for these operations can be further reduced by employing different architectural and circuit-level solutions. \textbf{Consequently, the focus of this work is to show the design space exploration for the NIST PQC finalist SABER with a focus on improving performance.}

The design space exploration, in this work, determines the adaption in various architectural elements (i.e., distinct memory configurations, pipelining, and logic sharing) with an emphasis on optimizing the design for a specific 65nm ASIC technology. Therefore, to initiate our design space exploration, we have selected an open source implementation of SABER\footnote{The utilized SABER core is modelled as an instruction set coprocessor architecture. The code is written in Verilog at  Register Transfer Level (RTL). It can be accessed directly at \url{https://github.com/sujoyetc/SABER_HW}.}. The existing code targets an FPGA platform, whereas in our work we target an ASIC platform. Converting the code to ASIC is one of the contributions of our work, as well as the following:

\begin{itemize}
    \item Exploration of different types, numbers, and sizes of compiled memories in a `smart synthesis' fashion.
    \item Promoting logic sharing between SABER building blocks that require similar functionality.
    \item Pipelining of selected portions of the design, thus trading-off throughput for latency.
    \item Design of a tapeout-ready SABER core in a commercial 65nm CMOS technology, for which we provide a layout and power, area, and timing characteristics.
    \item Source codes for our many architectures\footnote{Available from \cite{Imran_saberchip_2021}.} 
\end{itemize}

The remainder of this paper is organized as follows: Section \ref{sec:background} provides the required mathematical background and discusses the baseline architecture for the SABER PQC KEM protocol. Our design space exploration is given in Section \ref{sec:design_space_explorations}. Implementation results and a comparison to the state of the art is provided in Section \ref{sec:results_and_comparisons}. Finally, Section \ref{sec:Conclusions} concludes the paper.

\section{Preliminaries}\label{sec:background}
This section presents the required mathematical background and a description of the chosen baseline architecture for SABER.

\emph{Symbols (or notations).} \label{subsubsec:symbols_and_notations}
The $p$ and $q$ are modulo powers of 2. Set of integers is presented with $\Z$. Then the ring of integers modulo $p$ and $q$ is $\Z_p$ and $\Z_q$, respectively. The ring of polynomials for an integer $N$ is presented with $R_p=\Z_p[x]/\langle x^N+1 \rangle $ and $R_q=\Z_q[x]/\langle x^N+1 \rangle $ where $N$ is a fixed power of 2. Vectors are shown in bold and lower case font (e.g., \textit{\textbf{a}}).

\emph{Security strength.} \label{subsubsec:SABER_background}
The security strength relies on the hardness of module Learning With Rounding (Mod-LWR) problem. Therefore, a Mod-LWR sample is defined as follows: 

\begin{equation}\label{eq:Mod_LWR}
    \textit{$(\textbf{a}, b=\lfloor\frac{p}{q}(\textbf{a}^T \textbf{s})\rceil) \,  \in R_q^{l\times 1} \times R_p$}
\end{equation}

In Eq. \ref{eq:Mod_LWR}, \textit{\textbf{a}} is a vector of randomly generated polynomials in $R_q$, \textit{\textbf{s}} is a secret vector of polynomials in $R_q$ whose coefficients are sampled from binomial distribution, and the modulus $p < q$. The identification between Mod-LWR samples and uniformly random samples in $R_q^{l \times 1}\times R_p$ formulates the Mod-LWR problem. Therefore, this Mod-LWR problem is presumed to be computationally infeasible both on classical and quantum computers. Consequently, SABER is a good candidate for developing quantum-resistant cryptosystems.

\emph{PKE and KEM operations.} \label{subsubsec:operations}
SABER is a Chosen Ciphertext Attack, i.e., IND-CCA, secure KEM and Chosen Plaintext Attack, i.e., IND-CPA, secure public-key encryption (PKE) scheme. Therefore, the PKE crypto operations are the generation of pairs of public and private keys (PKE.KeyGen), encryption (PKE.Enc) and decryption (PKE.Dec). Similarly, the corresponding KEM operations are key generation  (KEM.KeyGen), encapsulation (KEM.Encaps) and decapsulation (KEM.Decaps). These operations are described as follows:

\textbf{Key Generation.} PKE.KeyGen starts by randomly generating a seed that defines an $l\times l$ matrix \textit{\textbf{A}} containing $l^2$ polynomials in $R_q$. A function $gen$ (see Algorithm 1 of \cite{Sinha_Roy_Basso_2020}) is used to generate the matrix from the seed based on SHAKE-128. A secret vector \textit{\textbf{s}} of polynomials is also generated. These polynomials are sampled from a centered binomial distribution. The generated public key contains a matrix seed and rounded product $\textit{\textbf{A}}^T\textit{\textbf{s}}$, while the secret key contains a secret vector $\textit{\textbf{s}}$. KEM.KeyGen does not differ from PKE.KeyGen, except that it appends a secret key with a hash of the public key and a randomly generated string $z$.

\textbf{Encryption and Encapsulation.} The PKE.Enc operation consists of generating a new secret $\textit{\textbf{s}}'$ and adding message to the inner product between the public key and the new secret $\textit{\textbf{s}}'$. This forms the first part of the
ciphertext while the second part contains the rounded
product $\textit{\textbf{As}}'$. The KEM.Encaps operation starts by randomly generating a message $m$ and obtaining from that the public key. The ciphertext $c$ contains the encrypted message and a value achieved from the message and public key.

\textbf{Decryption and Decapsulation.} PKE.Dec requires the secret key $\textit{\textbf{s}}$ to extract original message from the inner product between the public and secret keys. It is the reverse to PKE.Enc. KEM.Decaps re-encrypts the obtained message with the randomness associated with it and checks whether the ciphertext corresponds to the one received. 

\emph{Set of parameters.} \label{subsubsec:parameters}
For a security level equivalent to AES-128, AES-192, and AES-256, SABER provides three variants that are termed LightSABER, SABER, and FireSABER, respectively. All three variants use polynomial degree $N = 256$ and moduli $q = 2^{13}$ \& $p = 2^{10}$. They differ only in the module dimension, binomial distribution parameter ($\mu$), and the message space. For more details about security parameters, PKE and KEM operations, we refer readers to algorithms 1--6 of \cite{Sinha_Roy_Basso_2020}.

\subsection{Baseline architectures} \label{subsec:baseline_architecture}

\subsubsection{FPGA Coprocessor architecture of \cite{Sinha_Roy_Basso_2020}} \label{subsubsec:SOTA_processor}
As introduced in Section \ref{sec:introduction}, we have used an open source crypto core for which the target platform is FPGA. The coprocessor consists of: (i) a data memory (BRAM with a size of 1024$\times$64); (ii) a program memory; (iii) a dedicated finite state machine based (FSM) controller for orchestrating the SABER operations; and (iv) individual SABER building blocks. The building blocks include: (i) polynomial Vector-Vector multiplier wrapper; (ii) variants of secure hashing algorithms, i.e., SHA3-256, SHA3-512, and SHAKE-128; (iii) a binomial sampler; (iv) AddPack; (v) AddRound; (vi) Verify; (vii) Constant-time Move (CMOV); (viii) Unpack; (ix) CopyWords; and (x) BS2POLVEC\textsubscript{p}. 
 
A BRAM-implemented memory is used to keep initial, intermediate, and final results for the computation of required cryptographic operations. A program memory is employed to enable the coprocessor flexibility and its instruction set architecture (ISA) that comprehends a number of instructions required by (the variants of) SABER. For polynomial multiplication, inside the Vector-Vector multiplier, a centralized schoolbook multiplier architecture is utilized (described in \cite{Sujoy_DAC}). A sampler is required to compute a sample from pseudo-random input string for all KeyGen, Encaps, and Decaps operations. The verify block is responsible for comparing two byte strings of the same length. Based on the output of the verify unit, CMOV is responsible to either copy the decrypted session key or a pseudo random string at a specified memory location. The AddPack block computes coefficient-wise addition with a constant followed by generated message. Moreover, it packs the resultant bits into a byte string. Similarly, the AddRound block performs coefficient-wise addition of a constant followed by coefficient-wise rounding. The unpack unit converts a byte string into bit string. The BS2POLVEC\textsubscript{p} block converts the byte string into a polynomial vector. A dedicated FSM is responsible for interpreting incoming instructions from the program memory and to communicate/activate the individual building blocks.
 
\subsubsection{Our baseline architecture} \label{subsubsec:differences_SOTA_processor}
To achieve our design premise, i.e., high performance, we have constructed a baseline ASIC architecture for evaluation on a commercial 65nm technology. The first key difference with respect to \cite{Sinha_Roy_Basso_2020} is the replacement of the BRAM with an SRAM. The SRAM is generated by using a commercial memory compiler provided by a partner foundry. Initially, for the baseline architecture, the memory size is kept identical (1024$\times$64). We will later show many variants where the number of memory instances and their sizes are optimized with the aim of improving the clock frequency. 

It is important to note that our baseline architecture \textbf{remains a coprocessor architecture} and that the same ISA is utilized. We assume the program memory resides outside of the SABER accelerator core. The same building blocks utilized in \cite{Sinha_Roy_Basso_2020} are kept in our work, but most of them are modified during our optimizations, which we detail in the next section.

\section{Design Space Exploration Process}\label{sec:design_space_explorations}

\begin{figure}[]
 \centering
 \includegraphics[width=0.70\linewidth]{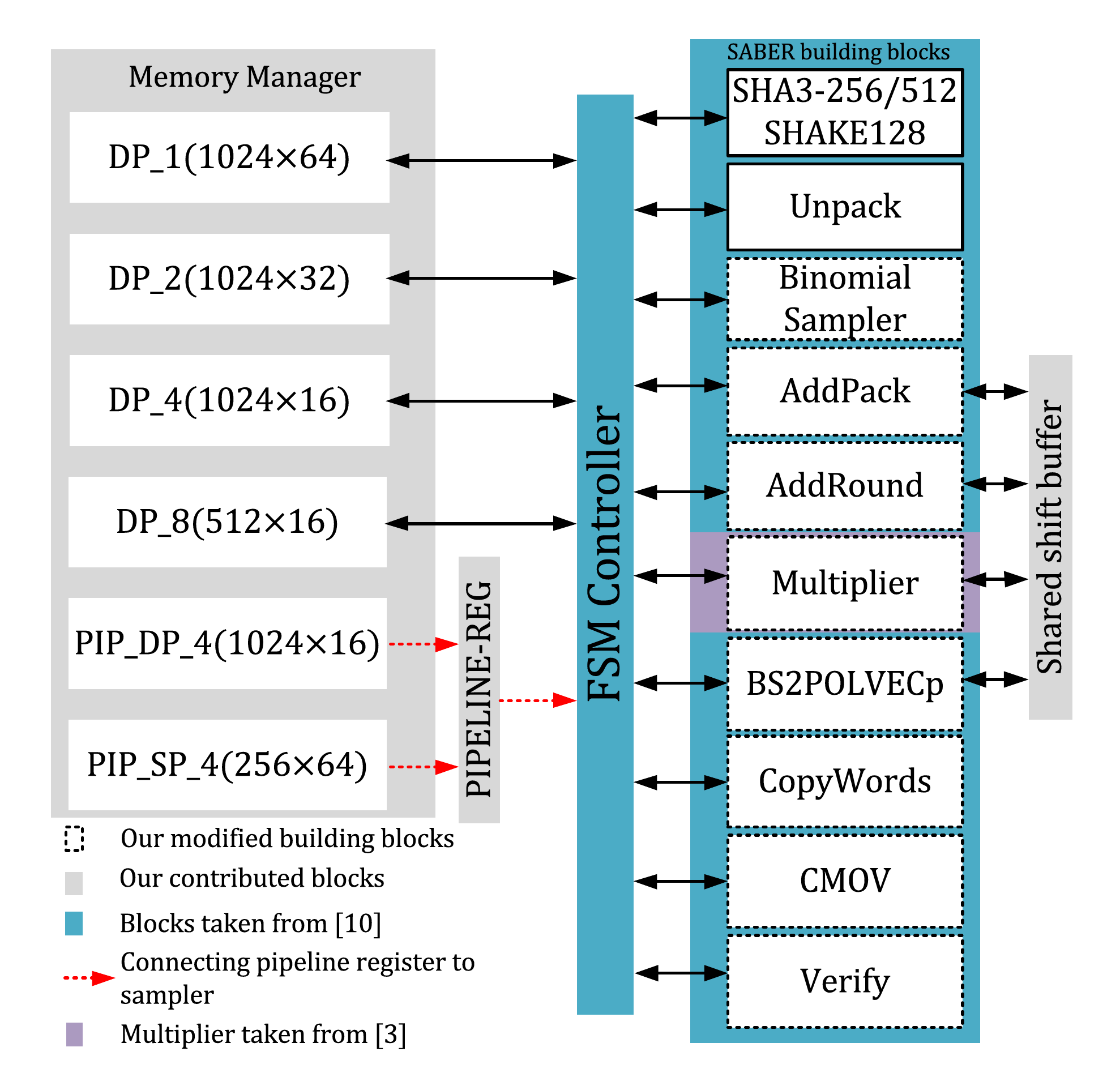}
   \vspace{-5pt}
 \caption{Block diagram of the designs generated during our design space exploration}
   \vspace{-5pt}

 \label{fig:block_diagram}
 \end{figure}

To differentiate our generated architecture to one another, we have adopted a different name for each design as shown in Fig. \ref{fig:block_diagram}. In order to provide a simple terminology for our studied architectures, we make use of the prefixes DP and SP, meaning that the architecture employs either a dual-port or a single-port memory. Similarly, the PIP prefix implies that the architecture in question is pipelined. Based on this terminology, the following architectures are considered:

\begin{itemize}
\item Baseline \hskip 9pt \{ \textbullet\enspace DP\_1(1024x64)
\item Optimized
  \(
  \left\{
    \begin{tabular}{@{\textbullet\enspace}l}
      DP\_2(1024x32) \\
      DP\_4(1024x16) \\
      DP\_8(512x16) \\
      PIP\_DP\_4(1024x16) \\
      PIP\_SP\_4(256x64)
    \end{tabular}
    \right.
  \)
\end{itemize}

Therefore, we have presented five optimized designs originating from our  baseline architecture. The memory is structured as \textbf{i(m $\times$ n)}, where $i$ is the number of instances, $m$ is the number of memory addresses, and $n$ is the data width of each address.


In addition to the FSM controller and building blocks shown in Fig. \ref{fig:block_diagram}, our design space exploration led to the creation of new units: (i) memory manager; (ii) pipeline register; and (iii) shared shift buffer. All these units are common to all of our studied architectures, except for the pipeline register that is employed only in our pipeline architectures, i.e., PIP\_DP and PIP\_SP. 
Furthermore, we have done modifications to many building blocks to synchronize their inputs/outputs with the memory timing requirements. The modified blocks are shown with dashed lines in Fig. \ref{fig:block_diagram}. 

\subsection{Memory manager}\label{subsec:memory_requirements}
A smart memory synthesis \cite{smartmem} approach is investigated and implemented in our Memory Manager unit. We clarify that the central concept of smart synthesis is the observation that having smaller and distributed memories can be advantageous in an ASIC design. Smaller memories require simpler address decoder units (which are faster). This, combined with the fact that part of the address decoding is now described as logic and can be co-optimized with the remainder of the design, leads to performance improvements with sometimes marginal increase in area. In this work, we explore a smart memory synthesis strategy within the limitations of a commercial memory compiler.

For KEM operations, when the security is equivalent to AES-192, SABER requires 992, 1344, and 1088 bytes for generating a single public-key, secret-key, and a cipher text \cite{SABER}. Therefore, a relatively large memory ($1024\times 64$) is employed in \cite{Sinha_Roy_Basso_2020}. We have used the same memory size in our baseline architecture. To initiate our design space exploration process, we have divided the data width (64 bit) of the employed memory into smaller chunks (32 and 16) and increased the number of memory instances accordingly. With this division, the memory structure becomes DP\_2(1024$\times$32) and DP\_4(1024$\times$16). This design choice results in an increase in clock frequency at the expense of area and power. Thereafter, from DP\_4(1024$\times$16) memory structure, we have constructed another architecture where we have reduced the required number of memory addresses from 1024 to 512. In this case, the memory structure becomes DP\_8(512$\times$16). Conversely, this design choice results in an increase in area and power with a marginal gain in clock frequency. Therefore, at this point, we deem that further diving the memories is no longer of interest.

In our first pipelined architecture, i.e., PIP\_DP, we have used the same 4(1024$\times$16) memory structure as employed in DP\_4(1024$\times$16). Our second pipelined architecture, however, utilizes compiled RegFiles\footnote{RegFiles are not flip-flops. This is a vendor-specific terminology for a compiled 6T SRAM memory that is advantageous when bit density can be traded-off with performance. It is also termed a ``high-speed'' variant of SRAM by its vendor.}. One of the limitations of the use of a RegFile is that the IP available to us is single-port, meaning that the design has to be modified such that all building blocks that benefit from concurrent read and write operations now execute them sequentially, one after the other. The consequence is that the overall number of clock cycles for a given cryptographic operation will increase. Later, we will show that this increase is beneficial since the improved clock frequency still reduces the overall latency for all SABER operations. The memory structure of the PIP\_SP architecture is 4(256$\times$64).


\begin{table*}[t]
    \caption{Logic Synthesis results for CCA-secure KEM SABER} \vspace{-5pt}
    \label{tab:implementation_results}
    \begin{tabular}{lllllllllll}
    \toprule
        \multirow{3}{*}{\textbf{Design}} & \multicolumn{2}{l}{\textbf{Area Information}} & \multicolumn{2}{l}{\textbf{Timing Information}} & \multicolumn{6}{l}{\textbf{Power Information (in \textit{\textbf{mW}})}} \\ \cline{2-11}
        {} & \multirow{2}{*}{Area ($mm^2$)} & \multirow{2}{*}{Gates} & \multirow{2}{*}{Clk. P ($ns$)} & \multirow{2}{*}{Freq. ($MHz$)} & \multicolumn{2}{c}{Crypto core} & \multicolumn{2}{c}{Combinational logic} & \multicolumn{2}{c}{Memory} \\ \cline{6-11}
        {} & {} & {} & {} & {} & Lkg & Dyn & Lkg & Dyn & Lkg & Dyn \\
    \midrule
        DP\_1(1024$\times$64) & 0.299 & 43336 & 2.000 & 500 & 0.090 & 86.844 & 0.059 & 16.235 (19\%) & 0.003 & 38.001 (44\%) \\
        DP\_2(1024$\times$32) & 0.308 & 45319 & 1.718 & 582 & 0.091 & 104.835 & 0.059 & 18.499 (18\%) & 0.004 & 48.322 (46\%)\\
        DP\_4(1024$\times$16) & 0.340 & 39981 & 1.638 & 610 & 0.082 & 135.342 & 0.051 & 18.762 (14\%) & 0.006 & 81.368 (60\%) \\
        DP\_8(512$\times$16) & 0.478 & 45979 & 1.624 & 615 & 0.099 & 220.410 & 0.062 & 21.691 (10\%) & 0.010 & 157.490 (71\%) \\
        PIP\_DP\_4(1024$\times$16) & 0.365 & 46217 & 1.508 & 663 & 0.097 & 233.361 & 0.063 & 20.890 (10\%) & 0.006 & 168.476 (72\%) \\
        PIP\_SP\_4(256$\times$64) & 0.314 & 64230 & 0.998 & 1002 & 0.111 & 142.413 & 0.074 & 32.925 (23\%) & 0.006 & 39.060 (27\%) \\
        \bottomrule
    \multicolumn{11}{l}{\textbf{Clk. P.} clock period, \textbf{Lkg.} leakage power, \textbf{Dyn.} dynamic power}
    \end{tabular}
\end{table*}

\subsection{Pipelining}\label{subsec:pipeline_register}
Initially, with the goal of improving clock frequency, we have employed different memory configurations until the improvements in clock frequency were exhausted. However, as the memory configurations change, the critical path of the design changes as well. In order to shorten the critical path and to further optimize the clock frequency, we have to explore other circuit level solutions, such as selective pipelining.

Based on the evaluation of the critical path of several architectures (details are given in section \ref{subsubsec:critical_path}), it becomes evident that the memory is the performance bottleneck of the design. For this reason, we have placed pipeline registers at the memory output. This guarantees that the critical path is proportional to the memory access time (as opposed to being proportional to the memory and to the logic that follows it). Therefore, in our PIP\_DP and PIP\_SP architectures, the input to the pipeline register is from the memory while the output is connected to the binomial sampler (not shown in Fig. \ref{fig:block_diagram}).

\subsection{Shared shift buffer}\label{subsec:shared_buffer}
For several building blocks of SABER, i.e., AddRound, AddPack, BS2POLVEC\textsubscript{p}, and multiplier, a shift register is required to read from many memory addresses and accumulate (hundreds of) bits into local registers. For example, a 320-bit long register is required in AddPack and BS2POLVEC\textsubscript{p} while a 64 and 676 bit register are required in AddPack and Multiplier, respectively. It is important to mention that all the SABER building blocks produce outputs serially, so the shift buffer can be shared as there are no concerns with concurrent access. Therefore, we have efficiently employed a single 676-bit register that is shared by AddRound, AddPack, BS2POLVEC\textsubscript{p}, and Multiplier. The use of a shared shift buffer results in a 10.3\% decrease in the total area with no impact on performance. All results given in the next section consider the use of this shared buffer by all architectures.

\section{Results and Comparisons} \label{sec:results_and_comparisons}
The synthesis results on a 65nm commercial technology for our baseline and optimized architectures are presented in Table \ref{tab:implementation_results}. These results are obtained after logic synthesis in Cadence Genus. The initial power estimates are obtained by assuming constant switching probabilities (i.e., while considering a synthetic workload). 


As shown in Table \ref{tab:implementation_results}, the concurrent use of compiled memories in a `smart synthesis' fashion with logic sharing to several SABER building blocks and pipelining allow us to achieve $1GHz$ clock frequency, albeit with overheads in area (column two) and power (columns six to eleven). With several optimizations from baseline (DP) to PIP\_DP architectures, we have shown that memory is the actual bottleneck in our implementation. For example, for baseline architecture, out of total dynamic power, the memory consumes 44\% while the combinational logic utilizes 19\%. Moreover, increase in memory instances results increase in power (72\% of the total dynamic power, see last column of Table \ref{tab:implementation_results} for our PIP\_DP\_4(1024$\times$16) architecture). Therefore, one approach to overcome this bottleneck is the use of faster memory instances as we employed in our PIP\_SP\_4(256$\times$64) architecture where combinational logic is responsible for 23\% of the dynamic power while memory is responsible for 27\%.

One interesting aspect of the PIP\_SP\_4 architecture is that the higher clock frequency changes the behavior of the synthesis tool considerably. We have verified that the tool then prefers to map the logic to (numerous) simpler gates instead of complex gates. Our analysis of the synthesis log also shows that partitioning decisions made by the tool were more frequent. The end result is that the PIP\_SP\_4 architecture has 18k more logic gates than its counterpart PIP\_DP\_4. We have also verified an increase in the number of buffers and inverters. Even for a simple gate like NAND2, we see 1626 instances in PIP\_DP\_4 while PIP\_SP\_4 has 3450 instances. It is important to highlight that the number of flip-flops does not change since the PIP\_SP\_4 design is identical to PIP\_DP\_4. 

We have calculated clock cycles (CCs) from end to end of each operation (KEM.KeyGen, KEM.Encaps, and KEM.Decaps). The time required to perform one cryptographic computation determines latency ($\mu s$) and is calculated using Eq. \ref{eq:latency}. The CCs information for each SABER building block is given in Table \ref{tab:CC_building_blocks}. The total CCs and latency to compute KEM.KeyGen, KEM.Encaps and KEM.Decaps for our baseline and optimized architectures is shown in Table \ref{tab:TCC_Latency}.

\vspace{-5pt}
\begin{equation}  
\label{eq:latency} 
    Latency\,(\mu s) = \frac{Total\,clock\,cycles}{Frequency\,(\textit{MHz})}
\end{equation}


\begin{table}[]
    \begin{threeparttable}
    \caption{CCs information for SABER building blocks} \vspace{-5pt} \label{tab:CC_building_blocks}
    \begin{tabular}{p{2.5cm}p{0.6cm}p{1.5cm}p{2.5cm}}
    \toprule
    \multirow{2}{*}{\textbf{building blocks}} & \multicolumn{2}{l}{\textbf{Clock cycles}} & \multirow{2}{*}{\textbf{Reason}} \\ \cline{2-3}
    {} & \cite{Sinha_Roy_Basso_2020} & This Work & {} \\
    \midrule
        Binomial Sampler & 145 & 246 & Pipelining \\
        Multiplier & 894 & 970 & Memory sync. \\
        Unpack & 167 & 295 & Memory sync. \\
        CopyWords & 60 & 211 & Single-port RegFile \\
        Others & - & \multicolumn{2}{l}{No change} \\
    \bottomrule
    \end{tabular}
    \end{threeparttable}
\end{table}


\begin{table}[]
    \begin{threeparttable}
    \caption{Total CCs and latency for CCA-secure KEM SABER on a 65nm commercial technology} \vspace{-5pt}\label{tab:TCC_Latency}
    \begin{tabular}{p{1.15cm}p{0.8cm}p{0.8cm}p{0.8cm}p{0.8cm}p{0.8cm}p{0.8cm}}
    \toprule
    \multirow{2}{*}{\textbf{Designs}} & \multicolumn{3}{l}{\textbf{Total clock cycles}} & \multicolumn{3}{l}{\textbf{Latency ($\mu s$)}} \\ \cline{2-7}
    {} & KeyGen & Encaps & Decaps & KeyGen & Encaps & Decaps \\
    \midrule
        {DP\_1} & {5644} & {6990} & {8664} & {11.2} & {13.9} & {17.3} \\
        {DP\_2} & {5644} & {6990} & {8664} & {9.6} & {12.0} & {14.8} \\
        {DP\_4} & {5644} & {6990} & {8664} & {9.2} & {11.4} & {14.2} \\
        {DP\_8} & {5644} & {6990} & {8664} & {9.1} & {11.3} & {14.0} \\
        {PIP\_DP\_4} & {5741} & {7087} & {8761} & {8.6} & {10.6} & {13.12} \\
        {PIP\_SP\_4} & {7154} & {7136} & {9359} & {7.1} & {7.1} & {9.3} \\
    \bottomrule
    \end{tabular}
    \end{threeparttable}
\end{table}

Table \ref{tab:CC_building_blocks} reveals that simultaneous use of multiple optimization approaches results in additional CCs when compared to baseline design. For example, our PIP\_SP\_4(256$\times$64) architecture requires 101, 76, 128, and 151 additional CCs for the Binomial Sampler, Vector-Vector Polynomial Multiplier, Unpack, and CopyWords building blocks. For other building blocks, the CC count will remain identical to the original design (meaning no changes when compared to \cite{Sinha_Roy_Basso_2020}). Similarly, Table \ref{tab:TCC_Latency} shows that the increase in both CCs and clock frequency (values given in column six of Table \ref{tab:implementation_results}) result in a decrease in the computation time. 

\subsection{Critical path analysis} \label{subsubsec:critical_path} 
\begin{figure}[]
 \centering
 \includegraphics[width=0.70\linewidth]{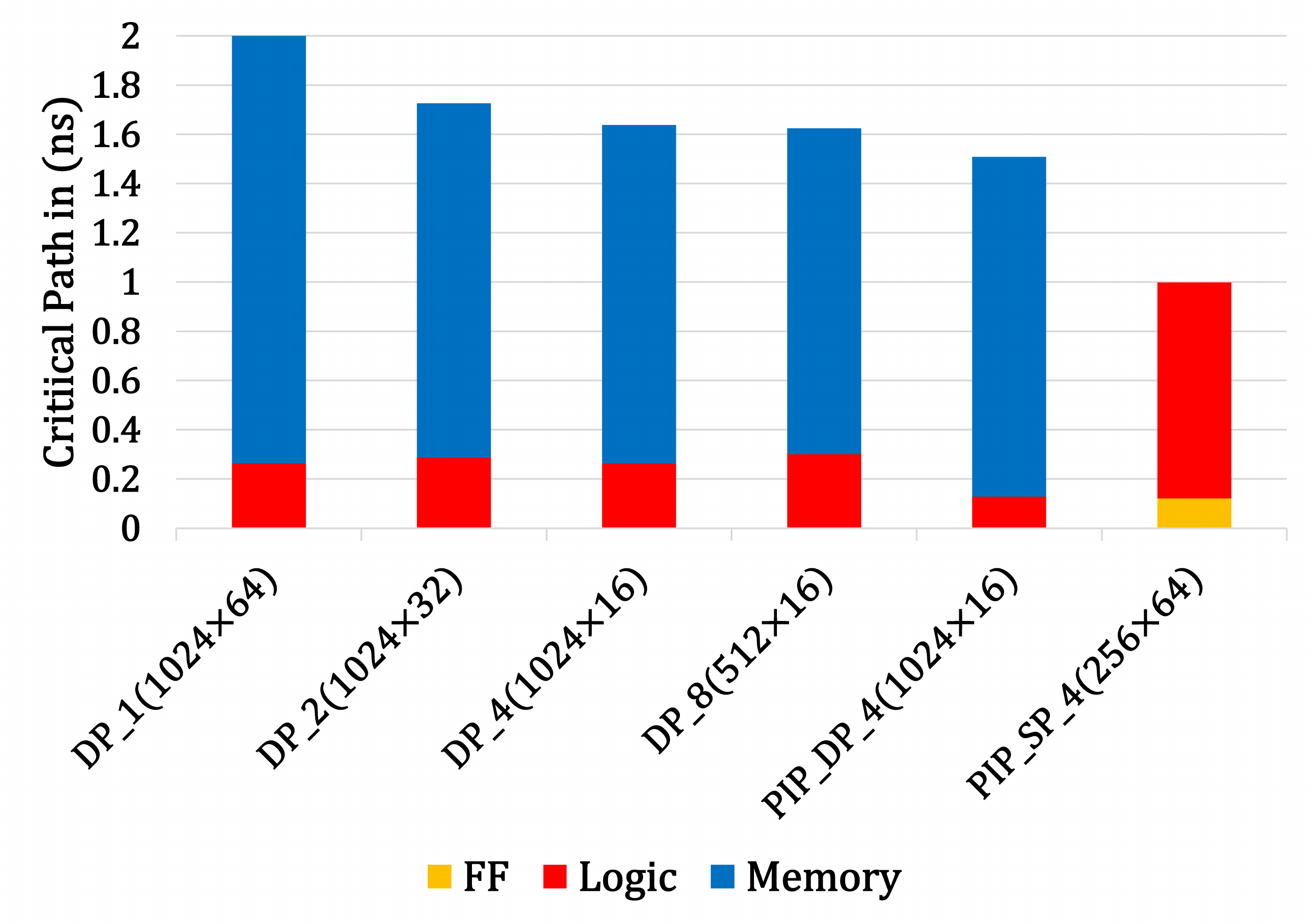}
\vspace{-10pt}
\caption{Critical path analysis for our studied architectures.}
\vspace{-10pt}
 \label{fig:critical_path}
 \end{figure}

The critical paths of our baseline and optimized architectures are shown in Fig. \ref{fig:critical_path}. Our analysis reveals that the memories containing longer access time result in longer critical paths for most architectures (i.e., the memory presents itself as the bottleneck) while the use of faster RegFiles result in a shorter critical path. In other words, as shown in Fig. \ref{fig:critical_path}, the critical path of our baseline architectures depend on the memory and some amount of combinational logic (to a lesser degree). However, this is not the case for our optimized PIP\_SP architecture where the critical path is mostly combinational logic (and the setup time of the destination flip-flop). This result implies that our optimized architecture is saturating the memory bandwidth thanks to our optimization strategies at architecture and circuit levels.


\subsection{Physical layout for PIP\_SP}\label{subsubsec:physical_layout}

\begin{figure}[]
 \centering
 \includegraphics[width=0.65\linewidth, trim={1cm 0.5cm 0.5cm 1cm} ,clip]{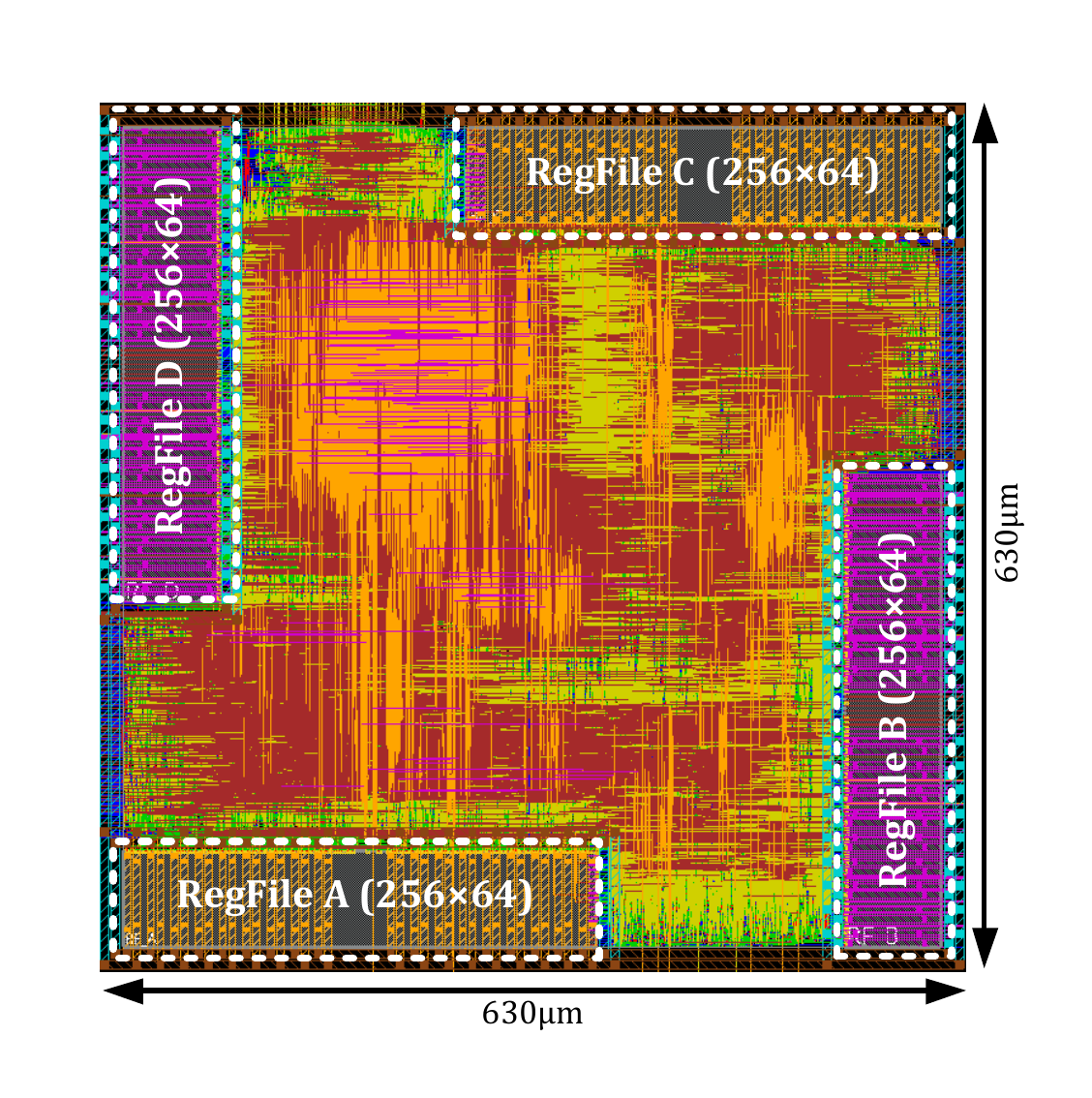} 
  \vspace{-10pt}
 \caption{Physical layout of the CCA-secure KEM SABER}
  \vspace{-10pt}
 \label{fig:physical_layout}
 \end{figure}

The layout of CCA-secure KEM SABER accelerator, as shown in Fig. \ref{fig:physical_layout}, is obtained from Cadence Innovus. The accelerator circuit was implemented with a nominal voltage of 1.2V in a 65nm CMOS technology. The design is placed and clock tree synthesis (CTS) is performed. The circuit is fully routed and passes design rule checking (DRC) with no violations. Metals M1 through M7 are used for signal routing, while the power is distributed in M8/M9. This is a typical metal stack for the considered 65nm process. The circuit is tapeout-ready with a core utilization of 88.66\%.


The results achieved after physical synthesis for different corners are given in Table \ref{tab:results_after_physical_synthesis}. These results were obtained with the aid of value change dump (VCD) files, i.e., files that capture the activity of the design based on representative simulation loads. Thus, the power values reported here are more realistic. Three different corners were used for characterization: slow-slow (SS), typical-typical (TT), and fast-fast (FF). These corners have operating conditions for different voltages and temperatures. The results reveal, as expected, that FF consumes more power than TT. Similary, TT consumes more power than SS. 
 
\begin{table}[]
    \begin{threeparttable}
    \caption{Power results for different process corners} \label{tab:results_after_physical_synthesis}
    \begin{tabular}{p{2.5cm}p{1.1cm}p{1.1cm}p{1.1cm}}
    \toprule
    \multirow{2}{*}{\textbf{Operations}} & \multicolumn{3}{c}{\textbf{Power values (in $mW$)}} \\ \cline{2-4} 
    {} & {\textbf{SS}} & {\textbf{TT}} & {\textbf{FF}} \\
    \midrule
    {KEM.KeyGen} & {146.7} & {184.3} & {244.8} \\
    {KEM.Encaps} & {148.9} & {187.0} & {248.3} \\
    {KEM.Decaps} & {148.4} & {186.4} & {247.5} \\
    \bottomrule
    \end{tabular}
    \end{threeparttable}
\end{table}

The comparison to existing SABER implementations is given in Table \ref{tab:comparisons}. Column one provides the reference implementation while the targeted platform is given in column two. The latency in $\mu s$ for KEM.KeyGen, KEM.Encaps and KEM.Decaps is given in column three. Column four provides the clock frequency ($MHz$). Finally, the last column provides the area for FPGA (in terms of look-up-tables and flip-flops) and ASIC (in $mm^2$) platforms. We have placed a `--' where required information is not available.


\begin{table}[]
    \begin{threeparttable}
    \caption{Comparison to existing SABER accelerators. All implementation results are for security equivalent to AES-192} \label{tab:comparisons}
    \begin{tabular}{p{0.85cm}p{1.45cm}p{1.7cm}p{0.6cm}p{2.2cm}}
    \toprule
    \multirow{2}{*}{\textbf{Ref. \#}} & \multirow{2}{*}{\textbf{FPGA/ASIC}} & \multirow{2}{*}{\textbf{Latency ($\mu s$)}} & \multirow{1}{*}{\textbf{Freq.}} & \multirow{1}{*}{\textbf{Area}} \\
    {} & {} & {} & {(\textit{MHz})} & LUT/FF (or) \textit{mm\textsuperscript{2}} \\
    \midrule
        {\cite{Lightweight_SABER}} & {Artix-7} & {--/467.1/527.6} & {100} & {6713/7363} \\
        {\cite{Three_Algos_2019}} & {Ultrascale+} & {--/60/65} & {322} & {--/--} \\
        {\cite{Roy_DAC_2021}}  & {Artix-7} & {3.2K/4.1K/3.8K} & {125} & {7.4K/7.3K} \\
        {\cite{Sinha_Roy_Basso_2020}} & {Ultrascale+} & {21.8/26.5/32.1} & {250} & {23.6K/9.8K} \\
        {\cite{SABER_Karatsuba_2020}} & {40nm} & {2.66/3.64/4.25} & {400} & {0.38} \\
        {\textbf{PIP\_SP}} & {65nm} & {7.1/7.1/9.3} & {1000} & {0.314} \\ 
    \bottomrule
    \end{tabular}
    \end{threeparttable}
    \vspace{-10pt}
\end{table}

\textbf{Comparison to FPGA implementations \cite{Three_Algos_2019,Roy_DAC_2021,Sinha_Roy_Basso_2020,Lightweight_SABER}}.
In terms of computation time (shown in Table \ref{tab:comparisons}), the most efficient implementation of SABER on FPGA is described in \cite{Sinha_Roy_Basso_2020}. It takes 5453, 6618 and 8034 CCs for the computation of one KEM.KeyGen, KEM.Encaps and KEM.Decaps which are comparatively 24\%, 8\% and 15\% lower than our PIP\_SP architecture. 
Moreover, our PIP\_SP architecture require 3.07, 3.73 and 3.45 times lower latency. For same operations, the proposed PIP\_SP architecture takes 450.7, 577.4 and 408.6 times lower latency as compared to \cite{Roy_DAC_2021}. Additionally, our PIP\_SP architecture achieves 8 and 4 times higher clock frequency as compared to \cite{Roy_DAC_2021} and \cite{Sinha_Roy_Basso_2020}, respectively. 

On Xilinx Zynq Ultrascale+ MPSoC, a software/hardware co-design processor architecture is presented in \cite{Three_Algos_2019}. For KEM.Encaps and KEM.Decaps, our PIP\_SP architecture is 8.45 and 6.98 times faster (in terms of latency). As compared to lightweight implementation of SABER, described in \cite{Lightweight_SABER}, our PIP\_SP architecture require 65.78 and 56.73 times lower latency for KEM.Encaps and KEM.Decaps, respectively. Moreover, our PIP\_SP architecture results 10 and 3.10 times higher clock frequency as compared to \cite{Lightweight_SABER} and \cite{Three_Algos_2019}. Noted that the area comparison to \cite{Lightweight_SABER,Sinha_Roy_Basso_2020,Roy_DAC_2021,Three_Algos_2019} is not possible due to distinct implementation platforms (as we have provided synthesis on ASIC while \cite{Sinha_Roy_Basso_2020,Roy_DAC_2021,Three_Algos_2019,Lightweight_SABER} utilizes FPGA). 

\textbf{{Comparison to ASIC accelerator \cite{SABER_Karatsuba_2020}}.}
As shown in Table \ref{tab:comparisons}, our optimized PIP\_SP architecture has higher latency. On the other hand, we are utilizing 1.21 times lower hardware resources on a 65nm technology while the referenced work utilized 40nm. It is therefore likely that our design would be a fraction of the size in the same technology. Moreover, we are achieving 2.5 times higher clock frequency. For multiplication of two 256-degree polynomials in SABER, we have employed a centralized schoolbook multiplier architecture of \cite{Sujoy_DAC}. It takes 256 CCs to compute one polynomial multiplication. On the other hand, in \cite{SABER_Karatsuba_2020}, the use of an 8-level Karatsuba multiplier for the same polynomial length requires 81 CCs instead of 256.

Furthermore, a high-speed Keccak module containing two parallel sponge functions (Keccak-f) is used in \cite{SABER_Karatsuba_2020}. It computes two Keccak-f[1600] computations in each clock cycle and each round of Keccak is performed every 12 CCs. In our architectures, a single sponge function in a serial fashion is incorporated which results in 28 CCs to generate 1,344 bits of a pseudo-random string. In addition to aforesaid differences in performance, our implementation follows a coprocessor architecture while a fully parallelized architecture is described in \cite{SABER_Karatsuba_2020}. Consequently, the decrease in clock cycles in \cite{SABER_Karatsuba_2020} ultimately shows decrease in computation time.

\section{Conclusions}\label{sec:Conclusions}
This work has presented a design space exploration of SABER with a focus on high performance. Our design space exploration results in $1GHz$ clock frequency with concurrent use of compiled memories in a `smart synthesis' fashion, logic sharing between SABER building blocks, and pipelining. Moreover, we have shown that for optimizing clock frequency with area and power overheads, a single instance of a large memory may not be optimal, and that numerous smaller memories can be more convenient. 

Finally, we highlight that our design already is tapeout-ready and will be sent for fabrication in early September (the packaged parts are expected to be delivered by December). This will allow us to extend this work with physical measurements after IC fabrication.

\section{Acknowledgments}
This work was partially supported by the EC through the European Social Fund in the context of the project ``ICT programme''. It was also partially supported by European Union's Horizon 2020 research and innovation programme under grant agreement No 952252 (SAFEST) and by the Estonian Research Council grant MOBERC35. 


\bibliographystyle{ACM-Reference-Format}
\bibliography{acmart}




\end{document}